\documentclass{jaa}
\usepackage[english]{babel}
\usepackage{epsfig}
\usepackage{epstopdf}
\usepackage{psfrag}
\usepackage{color}
\usepackage{graphicx}
\usepackage{fancyhdr}
\usepackage{graphics} 
\usepackage{amsmath}

\def\th{\vec{\theta}}
\def\u{\vec{U}}

\def\araa{ARAA}

\def\mnras{MNRAS}

\def\aap{A \& A}
\def\apj{Ap.J}

\def\u{{\bf U}} 

\def\V2{V_2}
\def\V2ij{V_{2ij}}
\def\S{{\mathcal S}}
\def\V{\mathcal{V}}
\def\N{{\mathcal N}}

\def\lsim{~\rlap{$<$}{\lower 1.0ex\hbox{$\sim$}}}

\def\gsim{~\rlap{$>$}{\lower 1.0ex\hbox{$\sim$}}}

\begin{document}

\title[Angular power spectrum of the Galactic synchrotron emission]
{The prospects of measuring the angular power spectrum of the diffuse Galactic
  synchrotron emission  with SKA1 Low}
\author[ S. S. Ali et al.] {Sk. Saiyad Ali$^{1}$\thanks{Email:saiyad@phys.jdvu.ac.in}, Somnath Bharadwaj$^{2}$, Samir Choudhuri$^{2}$, 
\newauthor Abhik Ghosh$^{3}$ and  Nirupam Roy $^{2,4}$ \\
$^{1}$ Department of Physics,Jadavpur University, Kolkata 700032, India\\  
$^{2}$ Department of Physics,  \& Centre for Theoretical Studies, IIT Kharagpur,  Pin: 721 302, India\\
$^{3}$ Kapteyn Astronomical Institute, PO Box 800, 9700 AV Groningen, The Netherlands\\
$^{4}$ Department of Physics, Indian Institute of Science, Bangalore 560012, India}
\date {}
\maketitle

%\maketitle
   
\begin{abstract}
The diffuse Galactic syncrotron emission (DGSE)  is the most important diffuse
foreground component for future cosmological 21-cm observations. The 
DGSE is also an important probe of the cosmic ray electron and magnetic field
distributions in the turbulent interstellar medium (ISM) of our Galaxy. In this paper 
we briefly review the Tapered Gridded Estimator (TGE) which can be used 
to quantify the angular power spectrum $C_{\ell}$ of the sky signal directly
from the  visibilities measured in radio-interferometric observations. 
The salient features of the TGE are (1.) it deals with the gridded data which
makes it computationally very fast (2.) it avoids a positive noise bias which
normally arises from the system noise inherent to the visibility data, and (3.)
it allows us to taper the sky response and thereby suppresses the contribution
from unsubtracted point sources in the outer parts and the sidelobes of the
antenna beam pattern.  We also summarize earlier work where the TGE was used to
measure the $C_{\ell}$ of the DGSE
using $150 \, {\rm MHz}$ GMRT data. Earlier measurements of  $C_{\ell}$ 
are restricted to $\ell \le \ell{max} \sim 10^3$ for the DGSE, the signal at
the larger $\ell$ values is dominated by the residual point sources after
source subtraction. The higher sensitivity of the upcoming SKA1 Low will allow
the point sources to be subtracted to a fainter level than possible with
existing telescopes. We predict that it will be possible to measure the
$C_{\ell}$ of the DGSE to larger values of $\ell_{max}$ with SKA1 Low. 
Our results show that it should be possible to achieve 
$\ell_{max}\sim 10^4$ and $\sim 10^5$ with $2$ minutes and $10$ hours of
observations respectively.
\\\\
{\bf{Key words:}} methods: statistical, data analysis - techniques: interferometric- cosmology: diffuse radiation
\end{abstract} 

%\begin{keywords}{methods: statistical, data analysis - techniques: interferometric- cosmology: diffuse radiation}
%\end{keywords}

\section{Introduction}
Observations of the redshifted 21-cm radiation from the large scale
distribution of neutral hydrogen (HI) have been perceived to be 
one of   the most promising probes to study the high redshift Universe
(Bharadwaj, Nath \& Sethi 2001; also recent reviews: Morales \& Wyithe 2010;
Pritchard \& Loeb 2012; Mellema et al. 2013). The principal challenge is
  to disentangle the cosmological 21-cm signal from the astrophysical foregrounds 
  which are 4-5 orders of magnitude brighter (Ali, Bharadwaj \&
  Chengalur 2008; Bernardi et al. 2009; Ghosh et al. 2012). Several methods of
foreground removal and foreground avoidance have been proposed in the
literature for detecting the Epoch of Reionization (EoR) 21-cm signal
(e.g. Chapman et al. 2014 and  references therein).

Theoretical modelling and  $150\, {\rm MHz}$ 
Giant Metrewave Radio Telescope
(GMRT{\footnote{http://www.gmrt.ncra.tifr.res.in}; Swarup et al. 1991)
observations show that 
extragalactic point  sources  are the most dominant  foreground component
at angular scales $\le 4^{\circ}$ (Ali, Bharadwaj \& Chengalur 2008;). These
are also   the angular scales relevant 
for telescopes like the Low-Frequency 
Array (LOFAR{\footnote{http://www.lofar.org/}; van Haarlem et al. 2013) and
the upcoming Square Kilometre Array
(SKA{\footnote{https://www.skatelescope.org}; Koopmans et 
 al. 2015). It is therefore clear that precise point   source subtraction is
 essential  for measuring the cosmological 21-cm signal. 
The residual data, after point source subtraction, is dominated by 
the  diffuse synchrotron emission from our Galaxy
(Bernardi et al. 2009; Ghosh et al. 2012).  There will also be a contribution
from the residual point sources below the flux limit of the data. 
These two residuals can themselves be be large enough to
 overwhelm the EoR 21-cm signal. Our present knowledge of the Galactic diffuse metre wavelength 
radio emission  is quite inadequate at  arcminute and sub-arcminute 
angular scales.   The mean
spectral index of the synchrotron emission at high Galactic latitude has been
constrained to be $\alpha = 2.5 \pm 0.1$ in the $100-200 \, {\rm
  MHz}$ range (Rogers \& Bowman 2008). The  angular power spectra of the diffuse
Galactic emission  has been measured in only a few fields (Bernardi et al. 2009; Ghosh
et al. 2012; Iacobelli et al. 2013).  Both the spectral index and the angular
power spectrum can vary significantly from field to field across different
directions in the sky, and much remains to be done towards quantifying
this. Further, the available measurements of the angular power spectrum of the
diffuse Galactic synchrotron radiation are restricted to $\ell \le 1,300$, the
signal at large $\ell$ being dominated by the residual point source
contamination. The upcoming SKA is expected to achieve considerably greater
sensitivity compared to the existing instruments, potentially leading to a
better quantification and understanding of the different foregrounds. In
particular, we can expect more accurate and deeper point source subtraction
which should allow us to probe the angular power spectrum of the Galactic
synchrotron emission  out to larger $\ell$ values (smaller angular scales)
than previously achieved. In addition to being an important foreground
component for the  cosmological  21-cm signal, 
the study of the diffuse Galactic synchrotron radiation is also important in
its own right. The angular fluctuations of the 
synchrotron radiation are directly related to the fluctuations in the magnetic
filed and also the fluctuations in the  electron density of the turbulent interstellar medium (ISM)
of our Galaxy (e.g. Waelkens, Schekochihin \& En{\ss}lin 2009; Lazarian \&
Pogosyan 2012;  Iacobelli et al. 2013),  a subject that is not very well
understood at present. In this paper we review the methodology for measuring
$C_{\ell}$ the angular power spectrum of the diffuse Galactic synchrotron
emission (hereafter, DGSE) using low frequency radio-interferometric observations, summarize
some of the existing observational results and then discuss the prospects for
the upcoming SKA.

In radio-interferometric observations, the quantity measured is the 
complex visibility. The measurement is done directly in Fourier
space which makes interferometers ideal instruments to quantify the
power spectrum of the sky signal. The visibility based power spectrum
estimators also have the added advantage that they avoid possible
imaging artifacts due to the dirty beam, etc (Trott et al. 2011).  A
visibility based estimator, namely the ``Bare Estimator'', has been 
successfully employed to study the power spectrum of the HI in the 
ISM of several nearby galaxies (eg. Begum et al. 2006;
Dutta et al. 2009) and also applied to  low frequency GMRT data
to  measure the angular power spectrum  
of the sky signal in   the context of cosmological HI observations   
(Ali, Bharadwaj \& Chengalur 2008; Ghosh et al. 2011a). The Bare Estimator
directly uses pairwise  correlations of the measured visibilities to estimate 
$C_{\ell}$, avoiding the self correlation that is responsible for a noise bias
in the estimator.  The Bare Estimator  
is very precise, but computationally very expensive for large data  
volumes, as the computation  scales as  the square of the  number of
visibilities (see   Figure 6 and 12 in Choudhuri et al. 2014). 
The Bare estimator has the added advantage that it does not pick up a noise bias
which is an issue for many of the other estimators.  For example, the image  
based estimator (Seljak 1997) used by  Bernardi et al. (2009) and  Iacobelli
et al. (2013) for $C_{\ell}$  
and the visibility based estimator (Liu \& Tegmark, 2012) for 
$P(k_{\perp},k_{\parallel})$ rely on modelling the noise properties of
the data and subtracting out the expected noise bias. However, the
actual noise in the observations could have baseline, frequency and
time dependent variations which are very difficult to model, and there
is always a possibility of residual noise bias contaminating the 21-cm 
signal. In a recent study, Paciga et al. (2011) have avoided 
the noise bias by cross-correlating the measured visibilities observed at
different times. This, however, implies that only a part of the available data
is actually being used to estimate the power spectrum. 

In this paper we highlight the Tapered Gridded Estimator (TGE) which 
is a novel estimator for measuring $C_{\ell}$ from the gridded visibility
data. The TGE is computationally fast and does not pick up a noise bias
contribution. It has the added feature of suppressing the 
contribution from the sidelobes and the outer regions of the antenna's filed of
view  through a tapering of the sky response (Ghosh etal. 2011b). 
In Section \ref{TGE} of this paper  we briefly summarize the theoretical
framework  (Choudhuri et al. 2014, 2016a,b) for the TGE. 
In Section \ref{sec:res} we review the earlier observational  work (Ghosh et al. 2012) 
which has used GMRT data  to quantify the angular power spectrum ($C_{\ell}$) of
the  DGSE at $150\, {\rm MHz}$.  Finally, we predict the contribution of two
major foregrounds namely the extra-galactic point sources and the DGSE, and   
make predictions for the prospects of measuring  the DGSE  using  future $160\,{\rm MHz}$
observations with the upcoming SKA1 Low.

\section{The Tapered Gridded Estimator}
\label{TGE}

In this section we  briefly review  the Tapered Gridded Estimator (TGE), the
details of which are presented in  Choudhuri et al. (2014).  In any
radio-interferometric observations the measured 
visibilities $\V(\u, \nu)$  are a sum of two components namely the sky signal
$\S(\u, \nu)$ and  the system noise $\N(\u,\nu)$ 
\begin{equation}
\V(\u, \nu)=\S(\u, \nu)+\N(\u,\nu) \,.
\label{eq:c1}
\end{equation}
The entire analysis here is restricted to a particular frequency channel $\nu$
and we do not show this explicitly in the subsequent discussion.

We assume that the signal and the noise are both uncorrelated Gaussian
random variables with zero mean. The measured visibilities record the
Fourier transform of the product of the primary beam pattern
${\mathcal A}(\th)$ and $\delta I(\th)$ the angular
fluctuation in the specific intensity of the sky signal. 
Here we use $\theta_{\rm FWHM}$  to denote the Full Width Half Maxima  (${\rm
  FWHM}$) of ${\mathcal A}(\th)$. The beam pattern ${\mathcal A}(\th)$ is well
modelled as a Gaussian ${\mathcal A}(\th)=e^{-\theta^2/\theta_0^2}$ with 
$\theta_0=0.6 \times \theta_{FWHM}$  in the central regions, the outer regions
and the sidelobes however are typically not so well quantified and the beam
pattern  also varies with time as the antennas rotate to track a source on the
sky.  The TGE
allows us to taper the sky response of the antenna elements and
thereby suppresses the contribution coming from the outer part of the
primary beam and the sidelobes. We implement the tapering by
multiplying the field of view with a frequency independent window
function, ${\cal W}(\theta)$. Equivalently, in the Fourier domain we
convolve the measured visibilities with $\tilde{w}(\u)$, the Fourier
transform of ${\cal W}(\theta)$. The convolved visibilities $\V_{cg}$ are
evaluated on a rectangular grid in $uv$ space using
\begin{equation}
\V_{cg} = \sum_{i}\tilde{w}(\u_g-\u_i) \, \V_i
\label{eq:1}
\end{equation}
where $\u_g$ refers to the different grid points and $\V_i$ refers to
the visibility measured  at a baseline $\u_i$.  We have chosen a  grid spacing
 $\Delta U=\sqrt{\ln2}/(2\pi\theta_w)$ which corresponds to one fourth of the FWHM of $\tilde{w}(\u)$ as an optimum value. Here, $\theta_w$ is the tapering window which is used in the frequency independent window function, ${\cal W}(\theta)$ (discussed later). For any fixed grid position $\u_g$, we have restricted the contribution to baselines $\u_i$  within $\mid \u_g - \u_i \mid \le 6 \Delta U$. The weight function  $\tilde{w}( \u_g - \u_i)$ falls considerably and we do not expect a significant contribution from the visibilities beyond this baseline. We note that the
gridding process considerably reduces the data volume and the
computation time required to estimate the power spectrum (Choudhuri et al., 2014).
It may be noted that tapering the sky response is
effective only if the window function $\tilde{w}(\u_g-\u_i)$ in
eq. (\ref{eq:1}) is densely sampled by the baseline $uv$ distribution. The GMRT  has a patchy $uv$ coverage and  the tapering may not be very
effective  in this case. The  width of the convolution window $\tilde{w}(\u_g-\u_i)$ increases  as the
value of tapering factor $f$ (discussed later) is reduced. The variation of the signal amplitude within the width of $\tilde{w}(\u_g-\u_i)$ becomes important at small baselines 
where it is reflected as an overestimate of the value of $C_{\ell}$ using TGE,  
though the excess  is largely within the $1\sigma$ errors. This deviation is
found to be reduced in a situation with a more uniform and denser baseline 
distribution like LOFAR  (see Figure 7 in Choudhuri et al. 2014 for details).
The problem of the sparse $uv$ coverage has been considerably reduced in an
improved version of the TGE (Choudhuri et al. 2016c), we however do not
sdiscuss this here.

\begin{figure}
\centering 
\psfrag{theta}[c][c][1.][0]{$\theta$ $[{\rm Degrees}$]}
\psfrag{Pbeam}[c][c][1.][0]{${\mathcal A}(\th)$}
\psfrag{Bessel}[r][r][0.7][0]{${\mathcal A}(\th)$}
\psfrag{taper3}[r][r][0.7][0]{${\mathcal A_W}(\th)$, $f=3.0$}
\psfrag{taper1}[r][r][0.7][0]{${\mathcal A_W}(\th)$, $f=1.0$}
\psfrag{taper0.7}[r][r][0.7][0]{${\mathcal A_W}(\th)$, $f=0.7$}
\includegraphics[scale=.80]{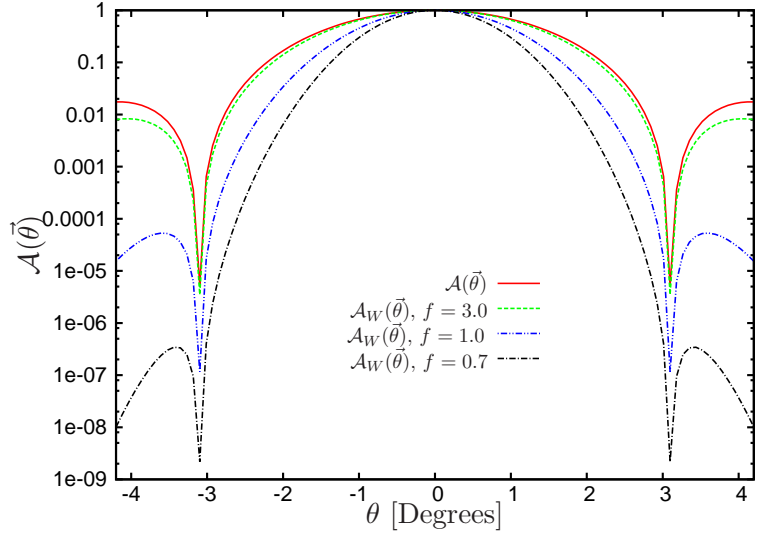}
\caption{The GMRT $150 \, {\rm MHz}$ primary beam ${\cal A}(\th)$
  which has been modelled as the square of a Bessel function.  The
  effective primary beam ${\cal A_W}(\th)$, obtained after tapering
  the sky response for different values of $f$ is also shown in the
  figure.}
\label{fig:taper}
\end{figure}

The convolved and gridded   visibility data $\V_{cg}$ 
gives  an estimate of the Fourier transform of the product of the intensity
fluctuations $\delta I(\th)$ and a modified primary beam pattern 
${\mathcal  A_W}(\th) ={\cal W}(\theta)\, {\cal A}(\th)$. For the purpose of
this paper we choose  a Gaussian window function ${\cal
  W}(\theta)=e^{-\theta^2/\theta^2_w}$ where we parametrize $\theta_w=f 
\theta_0$ and refer to $f$ as  
 the tapering factor which should preferably have a value  $f \le 1$ so that  
the window function cuts off the sky response well  before the first null of
${\mathcal A}(\th)$. We can approximate the modified primary beam pattern  as
a Gaussian  $ {\cal A_W}(\theta)=e^{-\theta^2/\theta_1^2}$ with $\theta_1=f
(1+f^2)^{-1/2} \theta_0$.  Figure \ref{fig:taper} shows ${\cal A}(\th)$ for
the GMRT, we have modelled this as a Bessel function corresponding to the
diffraction pattern of a circular  aperture of diameter $45 \, {\rm m}$. The
figure also shows the modified primary beam pattern ${\cal A_W}(\th)$ for the 
three different values  $f=3.0,1.0$ and   $0.7$. For $f=3.0$  we see that 
${\cal A_W}(\th)$ is almost same as ${\cal A}(\th)$ in the region   within
the first  null,  the difference however increases at the outer region of the
primary beam beyond the first null. We see that the effective primary beam
gets narrower as the 
value of $f$ is reduced.  The first side lobe of ${\cal A_W}(\th)$ is
suppressed by a factor of $\approx 10^{3}$ ($10^{6}$) compared to ${\cal
  A}(\th)$  at  $\mid \th \mid = 4^{\circ}$ for $f=1.0$ ($0.7$). We expect
that for $f = 0.7$ the  tapering will suppress by at least a factor of $10$  
the contribution from any residual point sources located beyond $2^{\circ}$
from the phase center.

 The correlation of the gridded visibilities $\langle \V_{c g} \V^{*}_{c g}
 \rangle$ provides a direct measurement of the angular power spectrum
 $C_{\ell_g}$, the angular brackets here denote the  ensemble average over
 many statistically independent realizations of the sky signal and the noise. 
In addition to the angular power spectrum $C_{\ell}$, the correlation 
 $\langle \V_{c g} \V^{*}_{c g} \rangle$ also picks up a contribution from 
the noise. This   introduces a positive noise bias if we use  $\langle \V_{c g}
\V^{*}_{c g} \rangle$ to estimate $C_{\ell}$. It is possible to eliminate this
positive noise bias by subtracting out the self correlation of the 
visibilities (see Section 5 of  Choudhuri et al. 2014 for details). 
We use this to  define the Tapered Gridded Estimator (TGE) at a grid point $g$ as 

\begin{equation}
{\hat E}_g=\frac{(\V_{cg} \V^{*}_{cg} - \sum_i \mid  \tilde{w}(\u_g-\u_i)
  \mid^2 \,   
\mid \V_i \mid^2)}{(\mid K_{1g} \mid^2 V_1 - K_{2gg} V_0)} \,,
\label{eq:2}
\end{equation}
 where the term $\sum_i \mid  \tilde{w}(\u_g-\u_i)  \mid^2 \mid \V_i \mid^2$
has been introduced  to cancel out the noise bias.  The denominator of 
eq. (\ref{eq:2}) is a normalization factor which converts the visibility
correlation to the angular power spectrum, and here 
 $V_0= \frac{\pi \theta_0^2}{2} \left( \frac{\partial
    B}{\partial T}\right)^{2}$, $V_1= \frac{\pi \theta_1^2}{2} \left(
  \frac{\partial B}{\partial T}\right)^{2}$, $K_{1g}=\sum_i
\tilde{w}(\u_g-\u_i)$, $K_{2g g}=\sum_i \mid \tilde{w}(\u_g-\u_i) \mid^2$, 
$B(T)$ is the Planck function and 
$\left(\frac{\partial B}{\partial T}\right)$ is the conversion factor from
temperature to specific intensity.

The estimator ${\hat E}_g$ defined here provides an unbiased estimate of
$C_{\ell_g}$ at the $\ell$ value $\ell_g=2 \pi \mid \u_g \mid$  corresponding
to the grid point $\u_g$ {\it i.e.}
\begin{equation}
\langle {\hat E}_g  \rangle =   C_{\ell_g}.
\label{eq:5}
\end{equation}
The $C_{\ell_g}$ values estimated at the different  grid points  are  binned  in
logarithmic intervals of $\ell$,  and we consider the bin-averaged $C_{\ell}$
as a function of the bin-averaged angular multipole $\ell$.  We have implemented the TGE in our earlier paper (Choudhuri et al. 2014) where we have assumed a uniform and dense baseline $uv$ coverage to calculate the normalization coefficient which relates visibility correlations to the estimated angular power spectrum $C_{\ell}$. There we have found that this leads to an overestimate of $C_{\ell}$  of the sky signal (diffues Galactic synchrotron emission) for instruments like the GMRT which have a sparse and patchy uv coverage. However, the  overestimation is  largely within the $1\sigma$ errors. This difference is found to be decreased in a situation with a more uniform and denser baseline distribution , like LOFAR.  We have reduced this problem due to sparse and patchy $uv$ sampling in an improved version of TGE (Choudhuri et al. 2016c) by calculating the normalization constant numerically using simulations.

 The analysis presented in the paper has been restricted to observations at a single frequency 
wherein the relevant issue is to quantify the two angular fluctuations (e.g. $C_{\ell}$) of the sky signal from the measured visibilities. This, however, is inadequate for the three dimensional redshifted HI 21-cm signal where it is necessary to also simultaneously quantify the fluctuations along the frequency direction. It may be noted that in a very
recent work (Choudhuri et al. 2016c) the TGE has been  further 
generalized to quantify the three dimensional power spectrum of the 21-cm brightness
temperature fluctuations $P(k_{\perp},k_{\parallel})$, we however do not discuss this here.

The TGE has three  key features: 

\noindent (a) It 
works directly with the measured visibilities  after {\it gridding} the data 
in the $uv$ plane. This significantly  reduces  the data volume and thereby
reduces the computation  required to estimate the power spectrum. This is
relevant  in the context of the current and the next generation
radio-interferometers which are expected to produce  large volumes
of visibility data in observations spanning many frequency channels and large
observing times.

\noindent (b) It is difficult to image and subtract point
sources from the outer parts and the sidelobes of the telescope's filed of
view. The residual point sources in the outer regions of the field of view  pose a
problem for estimating the power spectrum  of the diffuse 
radiation like  the Galactic synchrotron radiation or the cosmological 21-cm
radiation. The TGE  enables us to estimate the power spectrum  with a 
{\it tapered} sky response.  This suppresses the contribution from the
outer regions and the sidelobes of the telescope's beam pattern.

\noindent (c) The system noise inherent to the observations typically
introduces a positive noise bias in the estimated power spectrum. The TGE 
is devised in such a way that it internally computes the noise  bias 
and subtracts this out to  provide  an $unbiased$ estimate of the
power spectrum of the sky signal. 

\section{Results and conclusion}
\label{sec:res}
 In this section we briefly summarize and discuss the results originally presented
in Ghosh et al. 2012. Here the TGE was applied to the visibility data  
from GMRT $150 \, {\rm MHz}$ observations in four different fields referred to
as Fields I, II, III and IV respectively. The discussion here is restricted to
FIELD I which is the only field where  it was possible to detect the angular
power spectrum of the diffuse Galactic synchrotron radiation. The residual
point source contamination  was too high for the Galactic synchrotron
radiation to be detected in the three other fields. 

\begin{figure*}
\includegraphics[scale=0.8,angle=-90]{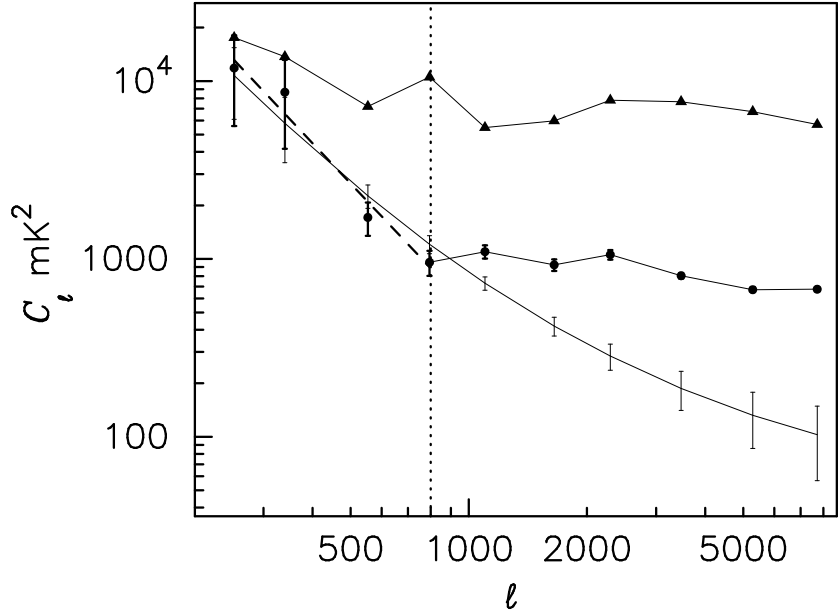}
\centering
\caption{This shows the measured $C_{\ell}$ before (triangle),  and after
(circle) point source subtraction. The $1\sigma$ error-bars are shown 
only for the residual data after subtracting   the sources above
  $S_C=10\,{\rm mJy}$. The residual  artifacts after point source subtraction 
have a peak flux $\sim  20 \, {\rm    mJy}$. The   dashed line  shows the best
fit  power-law  for $(\ell \le 800)$. The thin solid line
(bottom  curve)   with $1$ - $\sigma$ error-bars shows the model prediction of
Ali, Bharadwaj \& Chengalur (2008) assuming that all point sources above
$S_{c} = 20 \, {\rm    mJy}$ have been identified and removed from the
data. This figure is reproduced from Ghosh et al. (2012).} 
\label{clp}
\end{figure*}

 FIELD I was centered at sky position with ${\rm RA} \ = \ 05 \ {\rm h} \ 30 \
{\rm m} \ 00 \ {\rm s}$ and ${\rm Dec} \ = \ +60^{\circ} \ 00^{'} \ 00^{''}$
with a frequency bandwidth of $8 \, {\rm MHz}$ centered at $150 \, {\rm MHz}$,
however only a frequency width of $3.75 \, {\rm MHz}$ was used for the final
analysis. A continuum image of the field was first used to identify point
sources to a limiting flux of   $10 \, {\rm mJy}$. This  corresponds to 
$\sim 7$  times the off-source RMS. noise of $1.3 \, {\rm  mJy/beam}$ obtained
in the continuum image.  The point sources were subtracted from the visibility
data, and the residual visibilities were used for the subsequent analysis. 

The TGE was applied to the visibility data  to estimate the angular power spectrum 
$C_{\ell}$ both before and after the point sources were subtracted. The
analysis used the value $f=0.8$ for the tapering parameter. The results are
shown in  Figure \ref{clp} which has been reproduced from Ghosh et
al. 2012. It is clearly visible that subtraction of the point sources causes 
the sky signal to fall considerably at all angular multipoles. The $C_{\ell}$
for the residual data after point  source subtraction clearly shows 
different behaviour in two different ranges of $l$. 
 At low angular multipoles ($\ell \le800$), which correspond to angular scales
 larger than $10'$, we find a steep power law behavior which is typical of the
 Galactic synchrotron emission observed at higher frequencies and larger
 angular scales (e.g. Bennett et al. 2003). The angular power spectrum
 flattens out for $\ell > 800$, and we find that it remains nearly flat out to
 $\ell \approx 8,000$ which corresponds to angular scales of $\sim 1^{'}$.
The nearly flat region arises from the residual point sources which have a
flux below the threshold for source identification, and hence have not been
subtracted from the data. There is also a contribution from the residuals 
arising from {\bf errors} in modelling the brighter point sources. The residual
point source contribution  dominates  $C_{\ell}$ at small angular scales where
we are unable to make out the Galactic synchrotron radiation.

It is clear from this analysis that $C_{\ell}$, after point source
subtraction, is dominated by the diffuse Galactic synchrotron emission
at $\ell \le 800$ which corresponds to $\sim 10'$. We
have used a weighted least square to fit a power law model 

\begin{equation}
C^M_{\ell}=A\times \left(\frac{1000}{\ell} \right)^{\beta}
\label{eq:dgse}
\end{equation}

to the measured $C_{\ell}$  for $\ell \le 800$. We find the best fit  
values   $A=513\pm 41 \, {\rm mK}^2$ and $\beta=2.34 \pm 0.28$ for
which $C^M_{\ell}$  is also  shown in Figure \ref{clp}. Our findings  are
consistent with those of 
Bernardi et al. (2009)  who have analyzed $150 \, {\rm  MHz}$ WSRT
observations.  They have subtracted out point sources above 15 mJy,  
and used the resulting image to estimate the angular power spectrum $C_{\ell}$
which they find is well described by a power law for $\ell \le 900$. There has
also been considerable work on modelling the Galactic synchrotron radiation at
the higher frequencies relevant for the Cosmic microwave background
radiation (CMBR) (tens of GHz). These works predict a power law
behaviour $C_{\ell} \propto \ell^{-\beta}$ where $\beta$ has
values in the range $2.4$ to $3$ down to $\ell=900$ (Tegmark \& Efstathiou
1996; Tegmark et al. 2000).  Giardino et al. (2002) have analyzed the $2.4 \,
{\rm GHz}$ Parkes radio continuum  survey of total intensity of the southern
Galactic plane where they find a slope $\beta=2.37 \pm 0.21$ across 
the range $40 \le \ell \le 250$. Our slope, measured at smaller angular
scales, is consistent with these findings. 

As mentioned earlier, the DGSE is the most
important diffuse foreground for observations of the cosmological 21-cm
signal. The study of the DGSE is also important in its own right as it allows
us to probe the cosmic ray electron and magnetic filed distributions in the 
turbulent ISM of our Galaxy. We now discuss the prospects of measuring the
angular power spectrum $C_{\ell}$ of the DGSE using $160 \, {\rm MHz}$
observations with the  upcoming SKA1 Low. Present measurement of the $C_{\ell}$ of
the DGSE at $150 \, {\rm MHz}$ are restricted to $\ell \le \ell_{max}=900$,
the  
signal at larger angular multipoles is dominated by the residual point sources
after point source subtraction. It is anticipated that future observations
with SKA1 Low will allow the point sources to be identified and subtracted
down to a lower limiting flux $S_c$ than possible in the earlier observations
described here.  This should, in turn, allow us to measure the $C_{\ell}$ of the  
DGSE out to a larger value of $\ell_{\max}$ thereby allowing us to probe the
structure of the turbulent ISM at smaller angular scales than those possible 
in earlier observations. 

\begin{figure*}
\begin{center}
\psfrag{cl}[b][b][1.5][0]{$ C_{\ell} \,[\rm mK^2]$}
\psfrag{l}[c][c][1.5][0]{$\ell$}
\psfrag{lm}[c][c][1.5][0]{$\ell_{\rm max}$}
\psfrag{le}[c][c][.7][0]{$\ell_{\rm max}= 58000\,({\rm S}_{c}/10 \,\mu \rm Jy)^{-0.55}$}
\psfrag{sc}[c][c][1.5][0]{${\rm S}_c \,[\mu \rm Jy]$}
\centerline{{\includegraphics[scale=.50]{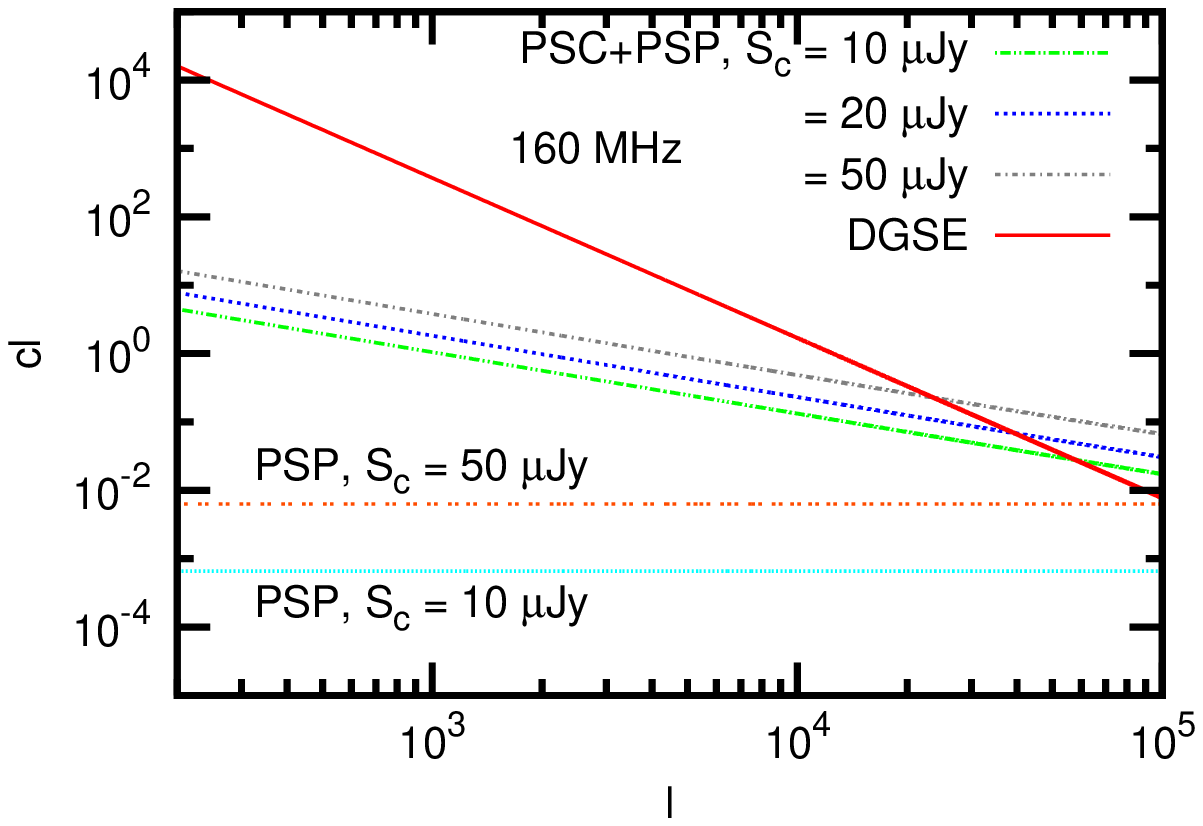}}
 {\includegraphics[scale=.50]{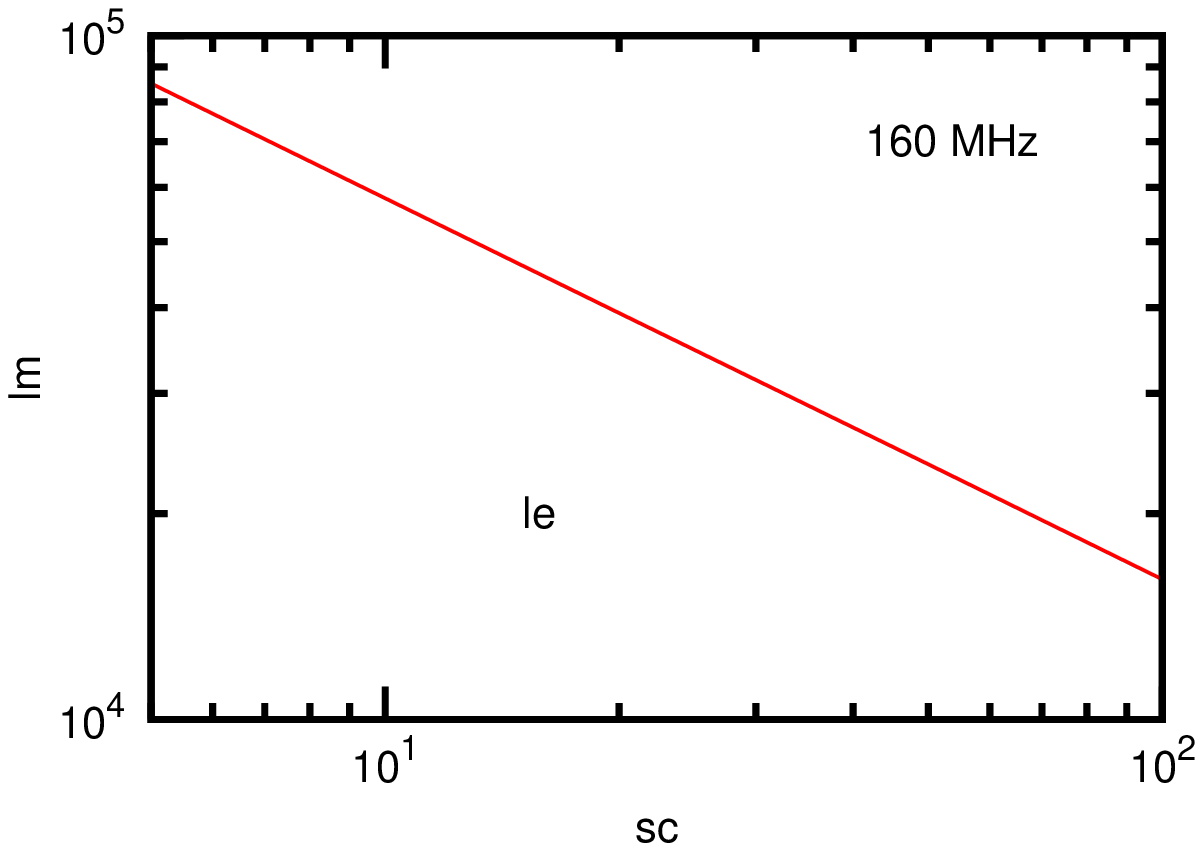}}}
\caption{{\bf{[Left Panel]}} Considering $160 \,{\rm MHz}$ SKA1 Low
  observations, this shows the model predictions for the angular power
  spectrum $C_{\ell}$ of the two major radio-sky components namely the
  DGSE and the point sources. The point source contribution itself is
  a combination of two components, the Point Source Clustering (PSC)
  and the Point Source Poisson (PSP) contributions respectively. The
  results for the point sources are shown for the different values of
  the limiting flux $S_c$ indicated in the figure. It is assumed that
  all the discrete sources brighter than $S_c$ have been identified in
  the image and subtracted from the visibility data.  {\bf{[Right
        Panel]}} This shows the maximum value of the angular multipole
  ($\ell_{max}$) below which ($\ell \le \ell_{max}$) we expect to
  measure $C_{\ell}$ for the DGSE, the signal is expected to be
  dominated by the residual point sources at $\ell > \ell_{max}$. The
  figure shows how the value of $\ell_{max}$ varies with $S_c$.  The
  value of $\ell_{max}$ is obtained by equating the total point source
  contribution (PSC+PSP) to the contribution from the DGSE (see the
  left panel). We find that the relation between $\ell_{max}$ and
  $S_c$ is well fitted by the power law shown in the figure.}
\label{fig:ps2nd}
\end{center}
\end{figure*}

 Table 6 of Prandoni and Seymour (2015) shows that one expects to  achieve 
a $5 \, \sigma$ noise level of $\sim 6\, \mu {\rm Jy}$/beam  in $\sim$ 8 hours
of  observation  in a single  field with SKA1 Low. We use this to estimate
the limiting flux $S_c$ for future SKA1 Low observations, assuming that all
sources above the $5 \, \sigma$ noise level can be identified and
subtracted. We further assume that the $5 \, \sigma$ noise level falls
inversely as the square-root of the observing time, and therefore the limiting
flux can  lowered by having longer observations. We use this to model the
angular power spectrum $C_{\ell}$ of the sky signal as the value of $S_c$ is
lowered (Figure \ref{fig:ps2nd}). Our prediction for $C_{\ell}$ incorporates
the two most dominant components, namely the point sources and the DGSE. 
The model prediction for the DGSE is based on the angular power spectrum
$C_{\ell}^M$ (eq. \ref{eq:dgse}) measured  in 150 MHz GMRT observation 
(Ghosh et al. 2012).  Point sources make two distinct contributions to the angular power
spectrum, the first being the Poisson fluctuation due to the discrete nature
of the sources and the second arising from the clustering of the sources. The
point source contribution is predicted using the differential source
count measured at 150 MHz (Ghosh et al. 2012) and a  power law index of  $1.1$
for the  angular two-point correlation $\omega(\theta)$ at $1.4\,{\rm GHz}$
(Cress et al. 1996). For the purpose of the predictions, the DGSE contribution 
has been extrapolate from $150\,{\rm MHz}$ to $160 \,{\rm MHz}$ using a mean
spectral index of 2.52 from  Rogers \& Bowman (2008). Similarly, we
have extrapolated the  point  source contribution to $160 \,{\rm MHz}$ using
a average spectral index of $2.7$ (Jackson 2005; Randall et al. 2012). The
framework for modelling the expected $C_{\ell}$ has been  presented in detail in Ali, Bharadwaj \&
  Chengalur (2008).

The left panel of Figure \ref {fig:ps2nd} shows the total point source
contribution to $C_{\ell}$  predicted for  three different values of the 
limiting  flux $S_c= 10 \, \mu {\rm Jy}$, $20 \, \mu {\rm Jy}$ and $50 \, \mu
{\rm Jy}$ respectively.  The two horizontal dotted lines  show just the
Poisson contribution   for $S_c= 10 \, \mu {\rm Jy}$ and $50 \, \mu {\rm Jy}$
respectively. 
The Poisson contribution falls very rapidly as $S_c$ is lowered, and we see
that the Poisson component  is sub-dominant to the clustered component 
at all values of $\ell$ for the $S_c$ values which we have considered here. 
The DGSE contribution shown in Figure \ref {fig:ps2nd} does not change with
$S_c$. We see that the DGSE is the most dominant component 
 at $\ell \le \ell_{max}=2\times 10^4$ and $\ell \le \ell_{max}= 5\times 10^4$
 for  $S_c=$ 50 $\mu {\rm    Jy}$ and 10 $\mu {\rm Jy}$ respectively.  
We clearly see that future observations with the  upcoming SKA1 Low are expected
to probe the DGSE at angular multipoles which are much larger than those where we
currently have measurements. 

 The right panel of Figure \ref {fig:ps2nd} shows how the value of $\ell_{max}$
 changes as the value of $S_c$ is varied. The largest angular multipole where
 we currently have measurements of the DGSE is around 
$\ell_{max} \sim 10^3$. 
We see that it is possible to achieve $\ell_{max} \sim 10^4$ , {\it
   i.e.} and order of magnitude increase in $\ell_{max}$,  with 
$S_c=100 \,\mu {\rm Jy}$ which one expects to achieve in approximately $2$
minutes of   observation with SKA1 Low. We find that the relation between 
$\ell_{max}$ and $S_c$ is well fitted by the power
law $\ell_{\rm max}= 58000\,({\rm S}_{c}/10 \,\mu \rm Jy)^{-0.55}$. This power
law implies  that we expect  ${\ell}_{max}$ to increase as $t^{0.275}$ as the
observing time $t$ is increased, and we can expect to achieve 
$\ell_{max} \sim 10^5$ in approximately $10$ hours of observation. 

In conclusion we note that present measurement of the angular power spectrum
$C_{\ell}$ of the diffuse Galactic synchrotron emission are restricted to
$\ell \le l_{max} \sim 10^3$. Future observations with the upcoming SKA1 Low
are expected to allow us to probe this at much larger multipoles. Our
predictions show that one expects to achieve $\ell_{max}$ in the range $10^4$
to $10^{5}$ using observations that span a few minutes to tens of hours in
duration.

\section*{Acknowledgment}
We thank an anonymous referee for useful comments and suggestions. SC would like to acknowledge the University Grant Commission (UGC), India for providing financial support through Senior Research Fellowship. SSA would like to acknowledge C.T.S, I.I.T. Kharagpur for the use of its facilities. SSA would also like to thank the authorities of the IUCAA, Pune, India for providing the Visiting Associateship programme.

\end{document}